\documentclass[aps,prl,twocolumn,showpacs,superscriptaddress,groupedaddress]{revtex4}

\usepackage{graphicx}
\usepackage{bm}        % for math
\usepackage{amssymb}   % for math
\begin{document}

\title{Distribution of counterions and interaction between two similarly charged dielectric slabs: Roles of charge discreteness and dielectric inhomogeneity }

\author{Weria Pezeshkian}

\affiliation{Department of Physics, Institute for Advanced Studies in
Basic Sciences (IASBS), Zanjan 45137-66731, Iran}

\author{Narges Nikoofard}

\affiliation{Department of Physics, Institute for Advanced Studies in
Basic Sciences (IASBS), Zanjan 45137-66731, Iran}

\author{Davood Norouzi}

\affiliation{Department of Biological Sciences, Institute for Advanced
Studies in Basic Sciences (IASBS), Zanjan 45137-66731, Iran}

\author{Farshid Mohammad-Rafiee}
\email{farshid@iasbs.ac.ir}

\affiliation{Department of Physics, Institute for Advanced Studies in
Basic Sciences (IASBS), Zanjan 45137-66731, Iran}

\affiliation{Department of Biological Sciences, Institute for Advanced
Studies in Basic Sciences (IASBS), Zanjan 45137-66731, Iran}

\author{Hossein Fazli}
\email{fazli@iasbs.ac.ir}

\affiliation{Department of Physics, Institute for Advanced Studies in
Basic Sciences (IASBS), Zanjan 45137-66731, Iran}

\affiliation{Department of Biological Sciences, Institute for Advanced
Studies in Basic Sciences (IASBS), Zanjan 45137-66731, Iran}

\date{\today}

\begin{abstract}
The distribution of counterions and the electrostatic interaction between two similarly charged dielectric slabs is studied in the strong coupling limit. Dielectric inhomogeneities and discreteness of charge on the slabs have been taken into account. It is found that the amount of dielectric constant difference between the slabs and the environment, and the discreteness of charge on the slabs have opposing effects on the equilibrium distribution of the counterions. At small inter-slab separations, increasing the amount of dielectric constant difference increases the tendency of the counterions toward the middle of the intersurface space between the slabs and the discreteness of charge pushes them to the surfaces of the slabs. In the limit of point charges, independent of the strength of dielectric inhomogeneity, counterions distribute near the surfaces of the slabs. The interaction between the slabs is attractive at low temperatures and its strength increases with the dielectric constant difference. At room temperature, the slabs may completely attract each other, reach to an equilibrium separation or have two equilibrium separations with a barrier in between, depending on the system parameters.

\end{abstract}

\maketitle

\section{Introduction}

Electrostatic interactions are of great importance in many biological and soft matter systems. For example, electrostatics plays a key role in proteins structure \cite{protein-folding}, interaction of DNA and proteins with charged ligands \cite{protein-DNA-ligand}, DNA packaging and condensation \cite{DNA-pack-condens}, conformation of polyelectrolytes in solution and their aggregation behavior \cite{PE,aggregation} and interaction between charged colloids and their collective behavior in solution \cite{colloids}. Numerous experimental and theoretical studies have been performed to understand these phenomena and to elucidate the role of electrostatic interactions. The study of the counterions distribution and electric potential around DNA \cite{DNA-study1}, the interaction between two DNAs \cite{DNA-study2} and the same studies on similarly or oppositely charged surfaces \cite{plate-study} are examples of these studies.

The difference between dielectric constant of water ($\simeq 80$) and that of organic materials ($\simeq 2$), silica ($\simeq 4$) and air ($\simeq 1$) is an important factor in electrostatic interactions of related systems. The role of such dielectric inhomogeneity has been investigated in systems like ionic channels \cite{ion-channel}, ions or colloids at the air-water interface \cite{ion-interface}, copolymers in electric field \cite{copolymer} and polyelectrolytes adsorbed to an interface \cite{PE-adsorption}. More efforts are needed to investigate the effects arisen from dielectric inhomogeneity on distribution of the counterions around charged objects and the interactions between them \cite{SC-dielectric1,SC-dielectric2}.

When the electrostatic interactions in a system are much stronger than the thermal energy, mean field theories such as Poisson-Boltzmann formalism are not suitable for describing the behavior of the system. In this situation, known as the strong coupling (SC) limit \cite{SC}, another theory, which is derived from the virial expansion of the system partition function, can be used. This theory has been used to study electrostatic interactions between charged objects and counterion distribution around them. Such studies have been performed for two parallel charged plates considering their surface charge density as uniform \cite{SC-uniform} and discrete \cite{SC-discrete1,SC-discrete2}. Also, electrostatic interaction between two overally neutral plates of randomly distributed annealed or quenched charges of both signs has been studied \cite{SC-random}. Corrections to the theory of Ref. \cite{SC} are suggested to describe the system behavior in a broader range of the parameters, namely the intermediate coupling regime \cite{theories}. Such corrections are recently formalized by expansion of the partition function around the ground state of the system in which the Wigner crystal is formed by the counterions \cite{Trizac-2011}. This approach gives a correction to the prefactor of the second term of the theory of Ref. \cite{SC} to agree with Monte Carlo simulations.  

%Recently, other theories have been introduced to cover the interval from weak to intermediate to strong coupling regimes \cite{theories}.

The system of two uniformly charged dielectric slabs has been studied taking into consideration the effects arising from dielectric inhomogeneities \cite{SC-dielectric1,SC-dielectric2}. It has been shown that in the SC limit the dielectric inhomogeneities cause accumulation of the counterions in the middle of the intersurface space between the slabs. Also, it has been shown in these studies that the dielectric inhomogeneities increase the repulsive pressure between the slabs at small inter-slab separations.

A widely used simplification in the studies of charged systems is the consideration of the charge distribution on the surfaces of charged objects as uniform. Clearly, this approximation is feasible only when the ions in the solution are far from the charged surfaces and also the separation between charged surfaces are quite larger than the typical separation between charged residues on the surfaces. When a point charge is very close to a big charged object or when two charged objects are very close to each other, discreteness of charge should be taken into account in calculation of the electrostatic interactions.
In the SC limit, it has been shown that the distribution of counterions in the vicinity of a discretely charged plate is very different from that of a uniformly charged plate. In the former case, counterions have a strong lateral correlation with the surface charges and counterion density on the surface is much higher relative to the case of uniformly charged surface \cite{SC-discrete1}. For the case of two discretely charged surfaces of the same sign, most of the counterions crowd near the surfaces and the attraction between the surfaces is stronger compared to the case of uniformly charged surfaces \cite{SC-discrete2}. The same differences have also been observed in the weak coupling limit \cite{WC-discrete1,WC-discrete2}.

The purpose of this paper is to consider the effects of both dielectric inhomogeneity and discreteness of charge on the electrostatic interaction between two similarly charged parallel dielectric slabs. 
We consider the charge distribution on the surface of each slab as square shaped uniformly charged regions of side length $l$ whose centers are on a square lattice of spacing $a$, as is shown in Fig. \ref{fig1}. The case of $l=a$ corresponds to uniform charge distribution on the slabs and $l\to 0$ corresponds to distribution of point charges. We use the Green function method  to calculate interaction energies of each counterion with its images and the surface charges, and use it to obtain the counterion density profile in the SC limit. Despite previous works (e.g. Refs. \cite{SC-discrete2,WC-discrete2}), all the Fourier modes of the charge distribution are taken into consideration here. This allows us to study the extreme limit of point charges on the surfaces, where the system behavior is noticeably different (see below). Interaction between the slabs is calculated by two approaches: The first approach is valid for low temperatures and the point-charge distribution on the slabs in which interactions of the counterions with each other is taken into account. The other approach is valid for room temperature and arbitrary values of $l$, which is accurate in the SC limit where the effect of counterions interactions on the pressure between the slabs can be ignored.

We find that consideration of the dielectric inhomogeneity in electrostatic interactions of the system increases the tendency of the counterions to gather in the middle of the intersurface space between the two slabs. Taking into account the discreteness of charge on the slabs however, increases the counterion density in the vicinity of the slabs surfaces. Resulted from these two competing effects, when the slabs are in far or close separations the counterions are mostly distributed in the vicinity of the surfaces or in the middle of intersurface space between them, respectively. In the limit of point-charge distribution of the surfaces, independent of the dielectric inhomogeneity strength, density of the counterions is maximum in the vicinity of the slabs surfaces at all inter-slab separations. It is found that the interaction between the slabs, at low temperatures is always attractive and its strength increases with increasing the difference between dielectric constants of the slabs and that of the environment. At room temperature, depending on the strength of charge discreteness and dielectric inhomogeneity, the system is found to have two different behaviors: the slabs stand in an equilibrium distance from each other or completely attract each other after passing a repulsive barrier in the intermediate separations. It is also found that for a range of the system parameters there are two equilibrium separations between the slabs corresponding to two minima in the free energy of the system. In the absence of dielectric inhomogeneity, by changing the charge distribution on the surfaces from uniform to point charges, the entropic repulsion barrier between the surfaces disappears and they attract each other at all values of inter-slab separation. These results are not observed in previous works and show the importance of considering the discreteness of charge. The effects of the dielectric inhomogeneity and the charge discreteness on the pressure between the slabs are also studied.

\section{The Model and the Green Function}

\begin{figure}[h]
\centering
\resizebox{0.99\columnwidth}{!}{
\includegraphics{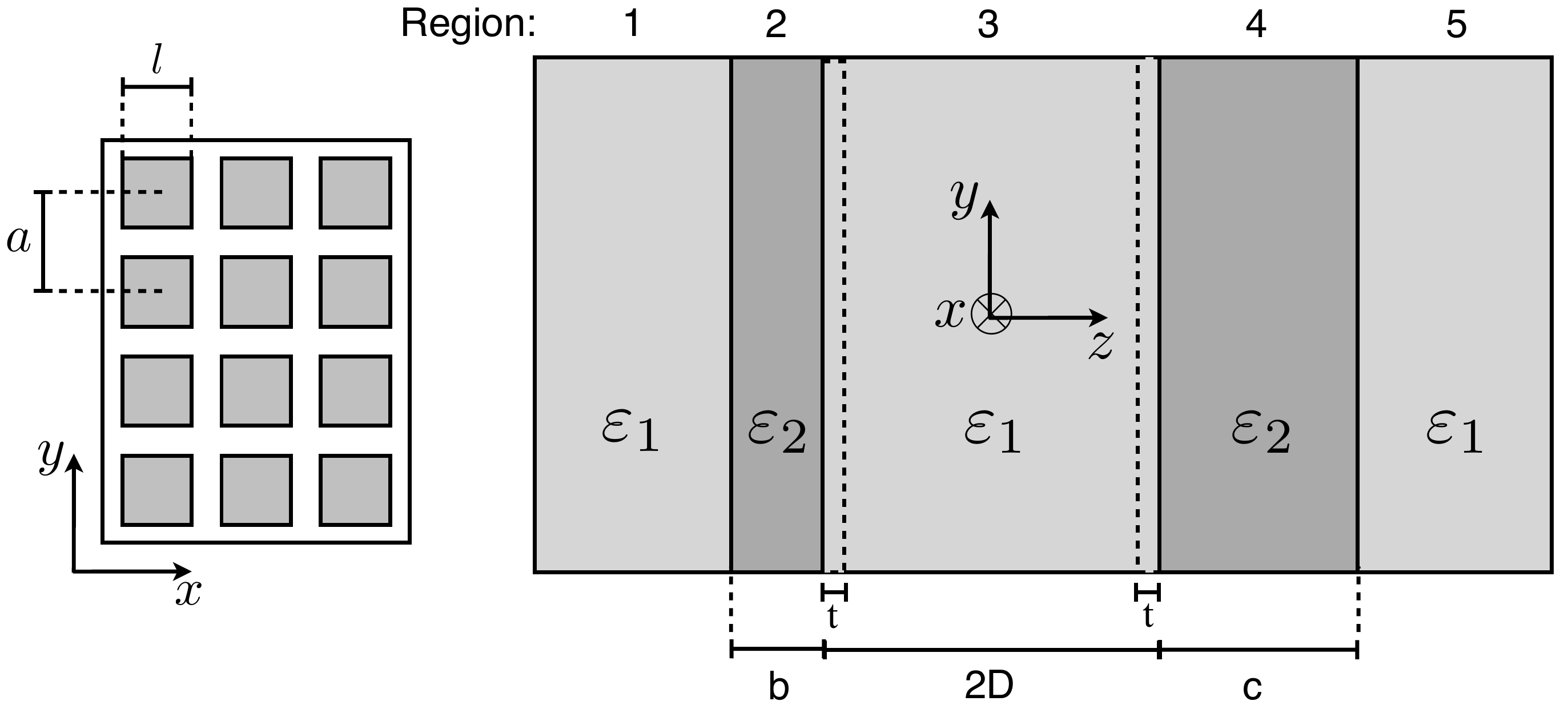}}
\caption{(Right panel) Two dielectric slabs of dielectric constant $\varepsilon_2$ placed in a medium of dielectric constant $\varepsilon_1$. The slabs are of thicknesses $b$ and $c$ and their distance is $2D$. The counterions are not permitted to enter a region of thickness $t$ on the slabs. (Left panel) On the inner surface of the slabs, charges are distributed on a lattice of constant $a$. On each lattice site, there is a square charged region of side $l$ and surface density $\sigma$.}
\label{fig1}
\end{figure}

We consider a system of two infinite dielectric slabs of thicknesses $b$ and $c$ with dielectric constant $\varepsilon_2$. The slabs are parallel and placed in a medium of dielectric constant $\varepsilon_1$ at separation $2D$ from each other (Fig. \ref{fig1}).
We assume that the charge on the inner surface of the slabs is distributed as a square lattice of square shaped regions of side $l$ and charge density $\sigma$. The centers of the square shaped regions form a square lattice of spacing $a$, as shown in Fig. \ref{fig1}. We define the cartesian coordinate system, in such a way that the $z$-axis is perpendicular to the surfaces and the origin is located in the middle of the separation area along the $z$ direction. By this definition, the position of the charged surfaces of the slabs along the $z$ direction is $z^* = \pm D$. The charge density of the slabs in three dimensions can be written as
\begin{eqnarray}
\rho(x,y,z) = \sigma\delta(z-z^*)\times {}
\nonumber\\
&&{}\hspace{-3.0cm}\sum_{m=-\infty}^{+\infty}\Theta(x+ma+l/2)\Theta(-x-ma+l/2) \times {}
\nonumber\\
&&{}\hspace{-3.0cm} \sum_{n=-\infty}^{+\infty}\Theta(y+na+l/2)\Theta(-y-na+l/2),
\end{eqnarray}
where $\delta(z)$ is the Delta function and $\Theta(x)$ is the theta function that is zero for $x<0$ and one otherwise. Keeping the value of $\sigma l^2$ fixed, $l\to 0$ corresponds to the distribution of point charges on a square lattice of spacing $a$ on the slabs surfaces, whereas $l=a$ corresponds to the uniformly charged surfaces of the slabs.  The case of $0<l<a$ describes the distribution of uniformly charged domains on the slabs surfaces. It is worth mentioning that to avoid any undesired singularities, a cut-off layer of thickness $t$ is supposed on each slab where the counterions cannot penetrate.

We consider the system in the SC limit. The coupling parameter is defined as $\Xi =\frac{q^2}{e^2} \frac{l_B}{\mu}$ in which $l_B=\frac{e^2}{4\pi\varepsilon_1\varepsilon_0k_BT}$ is the Bjerrum length outside the slabs, and $\mu=\frac{e^2}{2\pi ql_B\sigma_S}$ is the Gouy-Chapman length, where $\sigma_S$ denotes the surface charge density and is given by $\sigma_S \equiv (\sigma l^2)/a^2$. We note that $q$ is the counterions charge and $\sigma$ is the surface charge density of the charged regions.

To calculate the electrostatic potential and interaction energies of the system, we use the Green function method and solve the equation $\nabla^2G_i({\bf r},{\bf r}^{\prime})=-4\pi\delta\left({\bf r}-{\bf r^{\prime}}\right)\delta_{i3}$ with appropriate boundary conditions, $G_i=G_{i+1}$ and $\varepsilon_i\frac{\partial G_i}{\partial n}=\varepsilon_{i+1}\frac{\partial G_{i+1}}{\partial n}$, where $i=1, 2\dots,5$ represents the regions shown in Fig. \ref{fig1} and $\varepsilon_i$ is the dielectric constant of region $i$. By defining $\Delta \equiv \frac{\varepsilon_1-\varepsilon_2}{\varepsilon_1+\varepsilon_2}$, the Green function in the region between the slabs is found as
\begin{widetext}
\begin{eqnarray}\label{green}
G_{3}({\bf r},{\bf r}^{'})=\frac{1}{2\pi}\int_{-\infty}^{+\infty}
dk_{x}dk_{y}\frac{(e^{-kz_{>}}+W(kc)e^{-2kD+kz_{>}})(e^{kz_{<}}+W(kb)e^{-2kD-kz_{<}})}
{k(1-W(kc)W(kb)e^{-4kD})}e^{+ik_{x}(x-x')+ik_{y}(y-y')},
\end{eqnarray}
\end{widetext}
where $z_>$ ($z_<$) is the larger (smaller) of $z$ and $z^{\prime}$, and the function $W(x)$ is defined as
\begin{eqnarray}
W(x) \equiv \Delta \frac{1-e^{-2x}}{1-\Delta^2 e^{-2x}}. \label{W}
\end{eqnarray}
To find the Green function in the presence of a single dielectric slab, one should substitute $c=0$ in Eq. \ref{green}. Now the electrostatic energy of the system can be calculated numerically using the calculated Green function and all the Fourier modes of the charge distributions.

\subsection{Distribution of the counterions}

In the SC limit, distribution of the counterions can be found by calculation of the electrostatic energy of the system. Contribution of a counterion to the electrostatic energy of the system can be written as $E = U_{\Delta,int} + U_{slab}$, where $U_{\Delta,int}$ is the image-counterion interaction arising from the dielectric discontinuity across the interface, and $U_{slab}$ accounts for the counterion interaction with the physical charges on the slabs.  In order to find the image-counterion interaction energy of each counterion, we assume a point charge $q$ at position $\vec{r}_0=(x_0,y_0,z_0)$ located between the two slabs. One can find the potential at position $\vec{r}$ as $\phi_{\Delta}({\bf r})=\frac{q}{4\pi\varepsilon_2}G_3({\bf r},{\bf r_0})$, where the Green function is introduced in Eq. (\ref{green}).
Substituting this potential in $U=\frac{1}{2} \int \rho \left({\bf r}\right)\phi\left({\bf r}\right)dV$, the total electrostatic energy of the system in the presence of one counterion can be obtained. This energy also contains the self-energy of the point charge, which is infinite and independent of the slabs parameters. The self-energy can be obtained at the limit of $D \to \infty$. After subtracting the self-energy of the point charge, the total image-counterion electrostatic energy of a counterion between the two dielectric slabs is found as
\begin{widetext}
\begin{eqnarray}
U_{\Delta,int}=\frac{q^2}{8\pi\varepsilon_{2}} \int_{0}^{+\infty}dk
\frac{W(kc)e^{2kz_{0}}+W(kb)e^{-2kz_{0}}+2W(kc)W(kb)e^{-2kD}}{e^{+2kD}-W(kc)W(kb)e^{-2kD}}. \label{U-int}
\end{eqnarray}
\end{widetext}

The electrostatic interaction energy of a counterion with the charges of the slabs, $U_{slab}$, can be calculated using the relation $U = q \phi$. The potential of one dielectric slab containing square-shaped charged regions on its surface can be found using the Green function of Eq. (\ref{green}) by substituting $c=0$ as
\begin{widetext}
\begin{eqnarray}
\phi_{1slab} &=& \frac{8\sigma}{\varepsilon_{2}a^2}\sum_{m,n=1}^{+\infty}
\frac{e^{-kz}(W(kb)+1)}{kk_{x}k_{y}} \times  \sin(k_{x}l/2)\sin(k_{y}l/2)\cos(k_{x}x)\cos(k_{y}y)
\nonumber\\
&+&\frac{2\sigma l}{\varepsilon_{2}a^{2}}\sum_{n=1}^{+\infty}
\frac{e^{-k^{\prime}z}(W(k^{\prime}b)+1)}{k^{\prime 2}} \times \sin(k^{\prime}l/2)(\cos(k^{\prime}x)+\cos(k^{\prime}y))
\nonumber\\
&-& \frac{\sigma l^2}{2\varepsilon_{2}a^2}z, \label{phi-1slab}
\end{eqnarray}
\end{widetext}
where $k_x=\frac{2\pi n}{a}$, $k_y=\frac{2\pi m}{a}$, $k=\frac{2\pi}{a}\sqrt{n^2+m^2}$ and $k^{\prime}=\frac{2\pi n}{a}$. We denote the relative displacement of the two dielectric slabs containing square-shaped charged regions in the $x$ and $y$ directions by $\phi_x$ and $\phi_y$, respectively. Clearly, in addition to above mentioned parameters of the slabs, the interaction energy of the system depends on the values of $\phi_x$ and $\phi_y$ as well. Using the Green function of Eq. (\ref{green}), the potential of the two dielectric slabs in the space between them can be written as
\begin{widetext}
\begin{eqnarray} \label{eq:2slab-phi}
\phi_{2slab}=\frac{8\sigma}{\varepsilon_{2}a^2}\sum_{m,n=1}^{+\infty}
\frac{(e^{-kz+kD}+W(kb)e^{-kD+kz})(1+W(kb))}{kk_{x}k_{y}(e^{2kD}-W^2(kb)e^{-2kD})}  \sin(k_{x}l/2)\sin(k_{y}l/2)\cos(k_{x}x)\cos(k_{y}y)+ {} \nonumber \\
&& {}\hspace{-15.5cm}\frac{2l\sigma}{\varepsilon_{2}a^2}\sum_{n=1}^{+\infty}
\frac{(e^{-k'z+k'D}+W(k'b)e^{-k'D+k'z})(1+W(k'b))}{k'^2(e^{2k'D}-W^2(k'b)e^{-2k'D})} \sin(k'l/2)(\cos(k'_{x}x)+\cos(k'_{y}y))+{} \nonumber \\ && {}\hspace{-15.5cm}\frac{8\sigma}{\varepsilon_{2}a^2}\sum_{m,n=1}^{+\infty}
\frac{(e^{kz+kD}+W(kb)e^{-kD-kz})(1+W(kb))}{kk_{x}k_{y}(e^{2kD}-W^2(kb)e^{-2kD})} \sin(k_{x}l/2)\sin(k_{y}l/2)\cos(k_{x}(x+\phi_{x}))\cos(k_{y}(y+\phi_{y}))+ {} \nonumber \\
&& {}\hspace{-15.5cm}\frac{2l\sigma}{\varepsilon_{2}a^2}\sum_{n=1}^{+\infty}
\frac{(e^{k'z+k'D}+W(k'b)e^{-k'D-k'z})(1+W(k'b))}{k'^2(e^{2k'D}-W^2(k'b)e^{-2k'D})} \sin(k'l/2)(\cos(k'_{x}(x+\phi_{x}))+\cos(k'_{y}(y+\phi_{y}))-\frac{\sigma l^2}{\varepsilon_{2}a^2}D. \label{phi-2-slab}
\end{eqnarray}
\end{widetext}
The ion interaction energy with the charges on the slabs is obtained using this potential and the relation $U_{slab}=q \phi_{2slab}$.

In the SC limit, $\Xi \gg 1$, the counterions distribution in the system can be written as a virial expansion, and to the leading order one has \cite{SC}
\begin{equation} \label{leading}
\rho_{SC} = \alpha e^{-\frac{E}{k_BT}} + O\left( \Xi^{-1} \right),
\end{equation}
where $E = U_{\Delta,int} + U_{slab}$ is the total electrostatic energy of each counterion discussed above and $\alpha$ is a normalization prefactor that can be calculated from relation $\int\rho dV = N$, which yields the total number of the counterions in the system. In the following sections, we use this relation and the calculated energy of each counterion to find the counterion distribution in the presence of one or two dielectric slabs of discrete surface charge.

\subsubsection{One dielectric slab with discrete surface charge density }

\begin{figure}
\centering
\resizebox{0.9\columnwidth}{!}{
\includegraphics{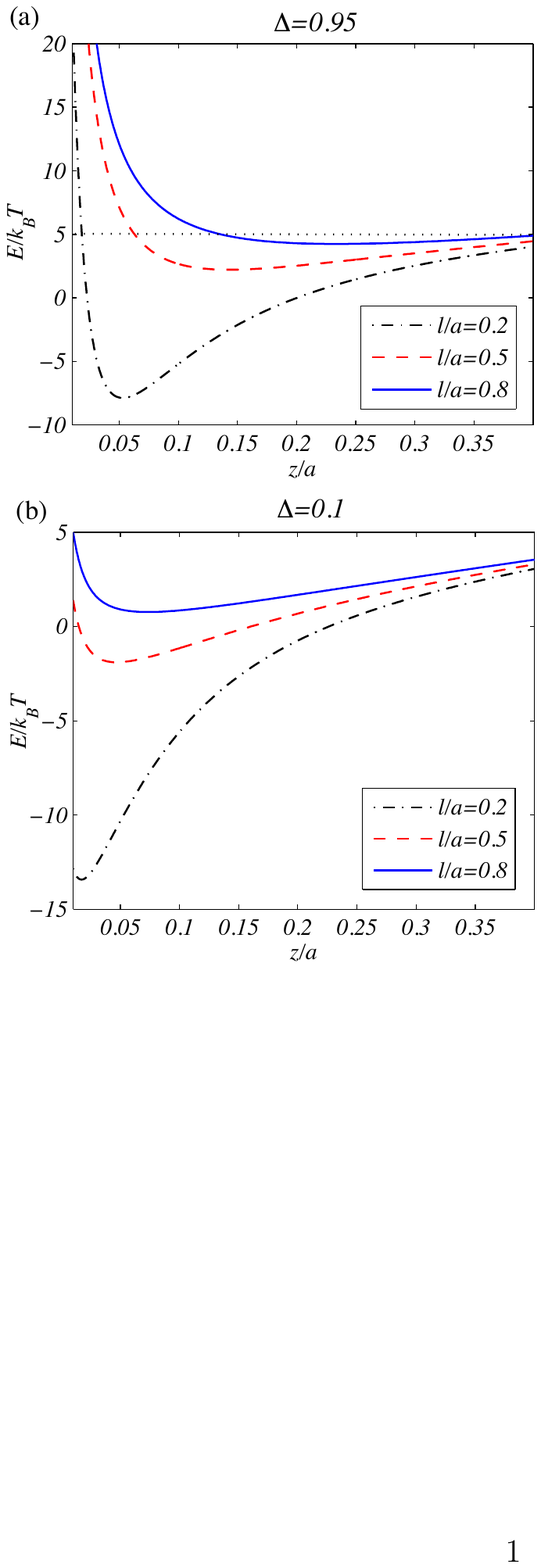}}
\caption{(color online) The total electrostatic energy per counterion for (a) $\Delta = 0.95$ and (b) $\Delta=0.1$. In both plots, the solid blue lines correspond to $l/a=0.8$, the red dashed lines correspond to $l/a=0.5$, and the black dashed-dotted lines correspond to $l/a=0.2$. }
\label{fig:1slab-energy}
\end{figure}

\begin{figure}
\centering
\resizebox{0.9\columnwidth}{!}{
\includegraphics{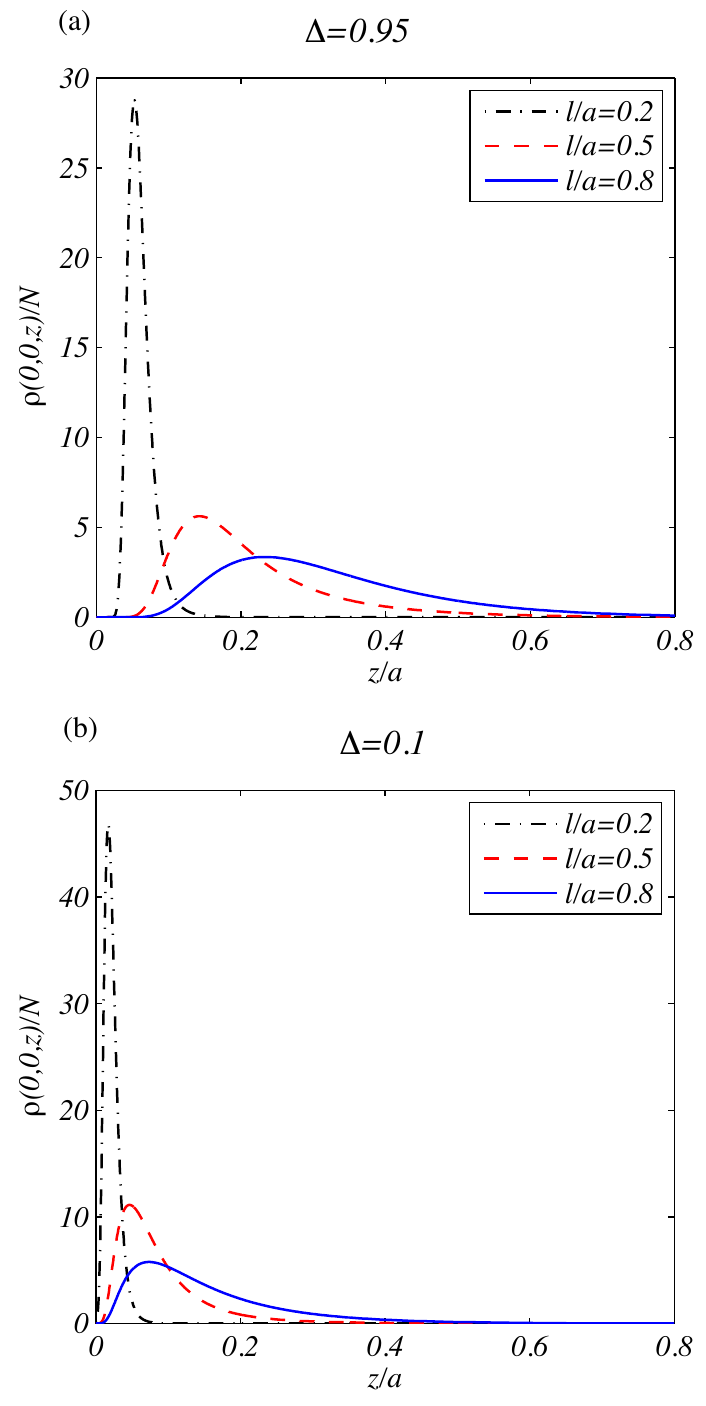}}
\caption{(color online) The density of the counterions along the $z$ direction for (a) $\Delta = 0.95$ and (b) $\Delta=0.1$. In both plots, the solid black lines correspond to $l/a=0.3$, the red dashed lines correspond to $l/a=0.4$, and the blue dashed-dotted lines correspond to $l/a=0.5$. }
\label{fig:1slab-density}
\end{figure}

\begin{figure}
\centering
\resizebox{0.9\columnwidth}{!}{
\includegraphics{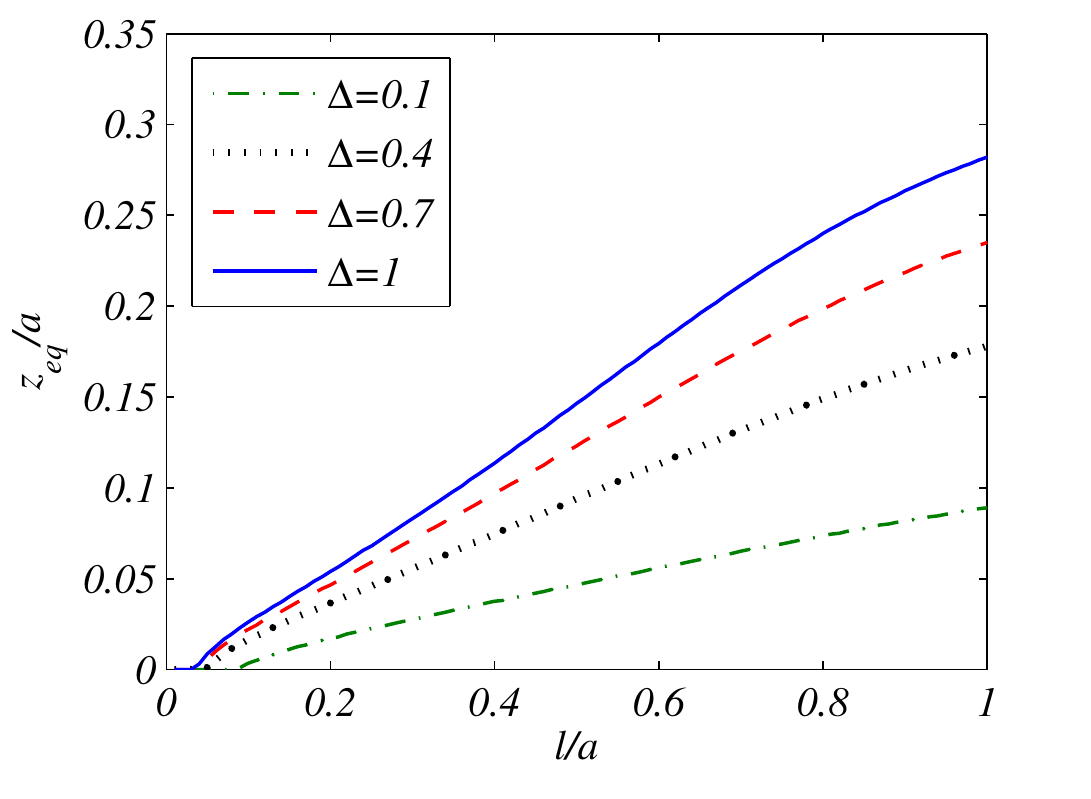}}
\caption{(color online) $z_{eq}$ as a function of $l/a$ for different values of $\Delta$. }
\label{fig:1slab-zeq}
\end{figure}

As mentioned above, the electrostatic energy of a counterion of charge $q$ in the presence of one dielectric slab can be written as
\begin{equation}
E = q\phi_{1slab}+\frac{q^2}{8\pi\varepsilon_{2}}\int_{0}^{+\infty}dkW(kb)e^{-2kz}, \label{E-1slab}
\end{equation}
where the first term is the electrostatic energy of the counterion with the physical charges on the slab and the second term shows the image-counterion interaction. To write the second term in Eq. (\ref{E-1slab}), we used Eq. (\ref{U-int}) and set the conditions of single slab, $c=0$ and $D=0$. The counterion interaction with the physical charges of the surface is attractive and long-ranged, whereas the image-counterion interaction is repulsive and short-ranged. For the numerical calculations and plots, we set $\sigma = e/l^2$ and $b = 2 a$. 
 We note that the counterions cannot enter a region of thickness $t$ on the slab. This thickness should be in the range of the molecular size, and we consider a representative value of $t=0.02 a$.
In Fig. \ref{fig:1slab-energy}, the behavior of energy, $E$, is shown as a function of $z$ for different values of $\Delta$ and $l$. In this figure, we set $x=y=0$. As it can be seen, the energy has a minimum in the $z$ direction, corresponding to a higher counterion density. In Fig. \ref{fig:1slab-density}, we present the profile of the counterion density in the $z$ direction for different values of $\Delta$ and $l$.  The counterion interaction with the physical charges of the slab is attractive, whereas the image-counterion interaction is repulsive. One can see that by decreasing $l$ (approaching to the point charge limit), the value of $z$ corresponding to the maximum of the counterion density shifts toward zero. Decreasing of $\Delta$ also pushes the counterions to the charged surface of the slab. At very small values of $l$, independent of $\Delta$, the counterions completely sit on the slab surface. This behavior is summarized in Fig. \ref{fig:1slab-zeq}, where $z_{eq}$ is shown as a function of the charge discreteness parameter, $l/a$. $z_{eq}$ is the position of the counterion in the $z$ direction, where the energy of Eq. (\ref{E-1slab}) is minimum.

\begin{figure}
 \centering
\resizebox{0.9\columnwidth}{!}{
\includegraphics{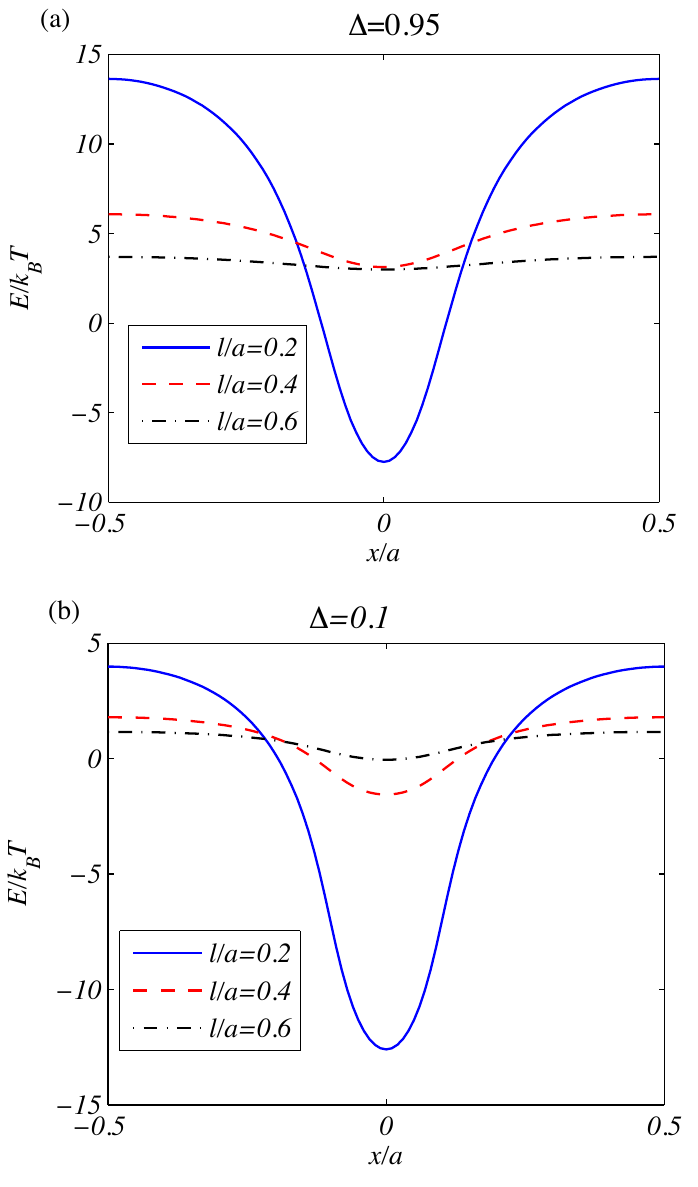}}
\caption{(Color online) Electrostatic energy per counterion as a function of $x/a$. The energy is calculated for $z=z_{eq}$ for each set of $l/a$. Plots (a) and (b) correspond to $\Delta=0.95$ and $\Delta = 0.1$, respectively.}
\label{fig:1slab-E-x}
\end{figure}

It is worth saying a few words about the lateral position of the counterion close to the slab. When the charge of the slab is discretized, the counterion prefers to be in front of the charged regions on the slab. In Fig. \ref{fig:1slab-E-x}, the energy of the counterion is shown in terms of $x/a$ for $z=z_{eq}$, $\Delta = 0.95$, and different values of $l/a$. As it can be seen, the minimum of the energy corresponds to $x/a=0$. At small values of $l/a$, the energy has a strong dependence on $x/a$ and the depth of the energy well in Fig. \ref{fig:1slab-E-x} relative to $k_BT$ is quite large. The dielectric inhomogeneity increases the depth of the energy well and the tendency of the counterions to stand in front of the charged regions of the slab.

\subsubsection{Two dielectric slabs with discrete surface charge density}

\begin{figure}
\centering
\resizebox{0.9\columnwidth}{!}{
\includegraphics{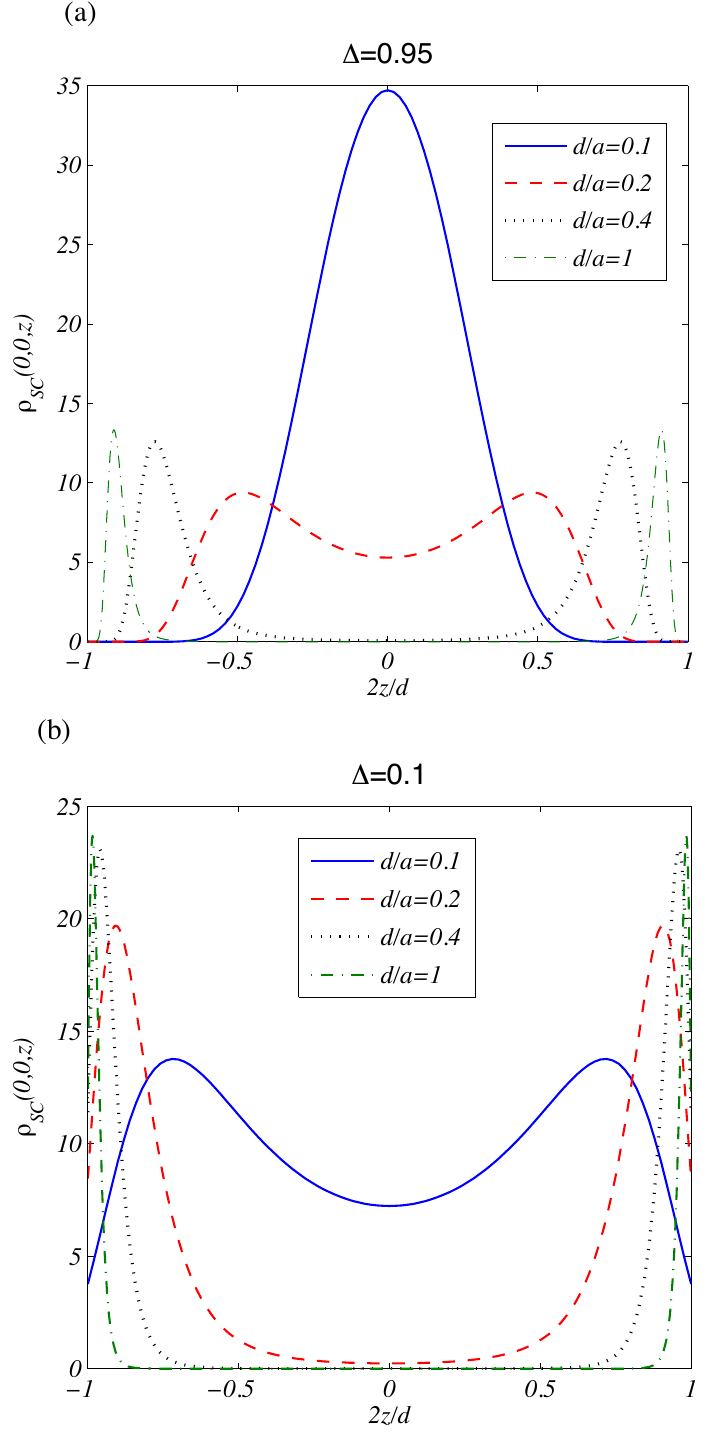}}
\caption{(Color online) The density of the counterions along the $z$ direction for $l/a=0.2$ and (a) $\Delta = 0.95$ and (b) $\Delta=0.1$. In both plots, the solid blue lines correspond to $d/a=0.1$, the red dashed lines correspond to $d/a=0.2$, the black dotted lines correspond to $d/a=0.4$, and the green dashed-dotted lines correspond to $d/a=1$.}
\label{fig:2slab-density-d}
\end{figure}

Similarly, we can calculate the electrostatic energy of a counterion of charge $q$ located between two dielectric slabs. To that end, we use Eq. \ref{eq:2slab-phi}, and find that the energy is minimum when the two slabs are completely in phase, meaning that $\phi_x=\phi_y=0$ (see Eq.  \ref{eq:2slab-phi}). The electrostatic energy of a counterion is found as
\begin{eqnarray} \label{eq:2slab-energy}
E &=& q\phi_{2slab}  \nonumber \\
&+& \frac{q^2}{4\pi\varepsilon}\int_{0}^{+\infty}dk\frac{W(kb)(\cosh(2kz)+W(kb)
e^{-2kD})}{e^{2kD}-W^{2}(kb)e^{-2kD}},
\end{eqnarray}
where the first term comes from the electrostatic interactions between the counterion and the physical charges on the two slabs, and the second term takes into account the image-counterion interactions. In this equation, it is assumed that the two slabs are of the same thickness $b$ and the interaction energy is found using Eq. \ref{U-int} and setting $c=b$. By defining $d \equiv 2(D-t)$ and using Eq. (\ref{leading}), one can find the counterion density profile between the two slabs. It is worth mentioning that it is possible to suggest several values for $t$ in the range of molecular size, and we consider a representative value of $t = 0.02 a$. Furthermore, in all the numerical calculations and the plots, we set $\sigma = e/l^2$ and $b= 2 a$. In Fig. \ref{fig:2slab-density-d}, the counterion density, $\rho_{SC}$, is plotted as a function of $z$ for different values of $\Delta$ and $d$. Interestingly, as it can be seen in this figure, when the two slabs are sufficiently far from each other (for example when $d \gtrsim 0.15 a$), the counterions prefer to be distributed in the vicinity of the slabs surfaces. This behavior means that the electrostatic interaction between a counterion and physical charges on the slabs is dominant. However, when the separation between the two slabs is small, the interaction between the counterion and the image charges dominates and the counterion moves to the middle of intersurface space between the slabs.

\begin{figure}
\centering
\resizebox{0.9\columnwidth}{!}{
\includegraphics{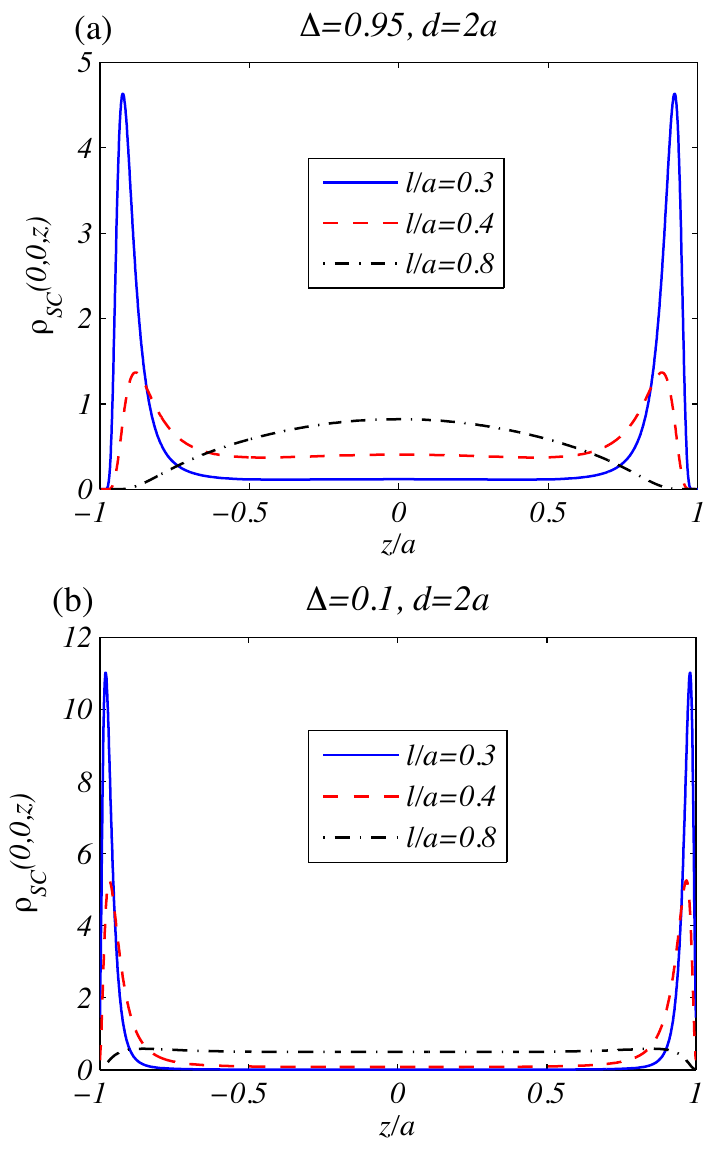}}
\caption{(Color online) The density of the counterions along the $z$ direction for $d/a=2$ and (a) $\Delta = 0.95$ and (b) $\Delta=0.1$. In both plots, the solid blue lines correspond to $l/a=0.3$, the red dashed lines correspond to $l/a=0.4$, and the black dashed-dotted lines correspond to $l/a=0.8$.}
\label{fig:2slab-density-l}
\end{figure}

\begin{figure}
\centering
\resizebox{0.9\columnwidth}{!}{
\includegraphics{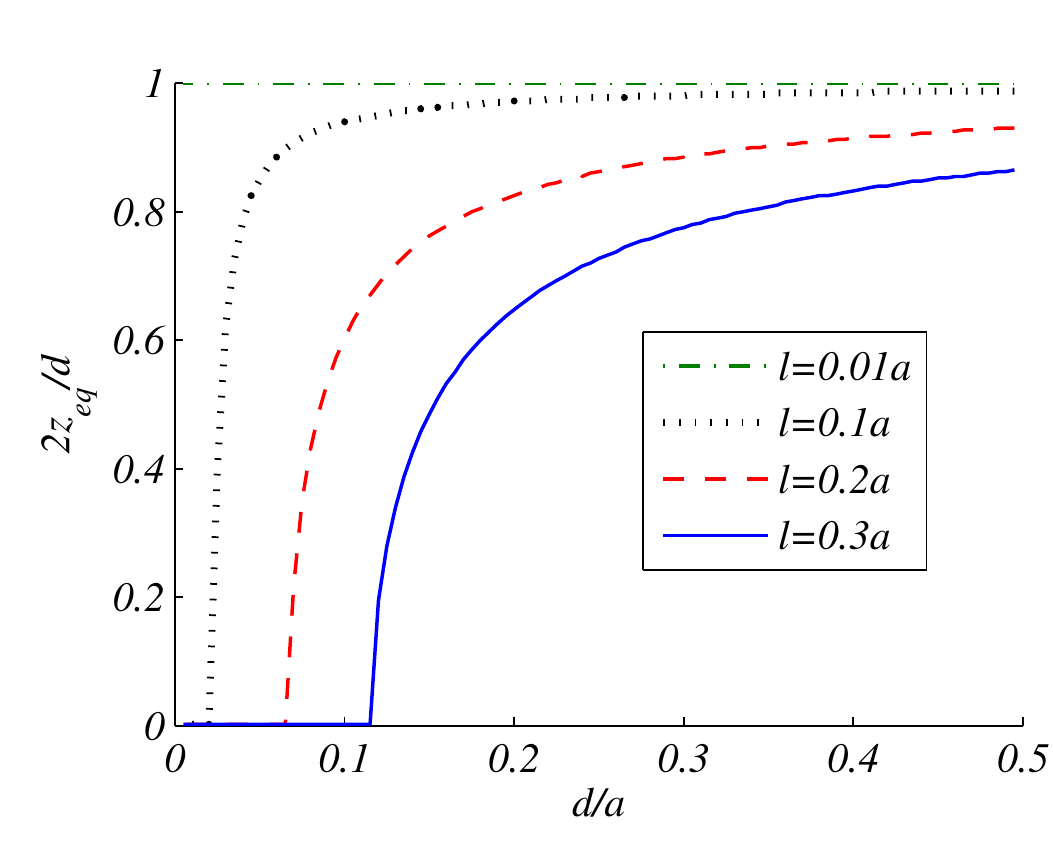}}
\caption{(Color online) $z_{eq}$ as a function of $d/a$ for $\Delta=0.95$ and different values of $l$.}
\label{fig:2slab-zeq}
\end{figure}

It is also interesting to study the effect of the discreteness of charge on the slabs. In Fig. \ref{fig:2slab-density-l} the dependence of $\rho_{SC}$ on the parameter of the charge discreteness, $l/a$, is shown for $\Delta=0.95$ and $d/a=2$. As it can be seen, the density of the counterions in the vicinity of the slabs increases with decreasing $l/a$. Furthermore, when the discreteness of charge on the slabs is washed out, the counterions prefer to be located in the middle of the space between the two slabs. To study the effects of discreteness of charge on the slabs and the distance between the two slabs on the preferred positions of the counterions, we define $z_{eq}$ as the preferred position of the counterion in the $z$ direction, which corresponds to the position that $\rho$ is maximum. In Fig. \ref{fig:2slab-zeq} the behavior of $z_{eq}/d$ is shown for $\Delta = 0.95$ and different values of $l/a$. The plots show that for sufficiently large values of $d$, $z_{eq}$ increases as $l/a$ decreases. In very small values of $l$, the counterions distribute near the surfaces of the slabs at all separations.

\subsection{Interaction between the two slabs}

In this section, we study the interaction between the two slabs in two conditions: The limit of $l \to 0$ and low temperatures, and the SC regime at finite temperatures. In the first approach despite the second one, interaction between the counterions is considered. In the SC regime, effect of the interaction between the counterions can be ignored. In this regime, the counterions freeze in a 2D lattice parallel to the surfaces, due to their strong interactions with each other relative to the thermal fluctuations \cite{SC}. The two approaches below are only applicable for divalent counterions, when counterions charge is twice the charge of each square-shaped region on the slabs ($q=2\sigma l^2$).

\subsubsection{Interaction between the two slabs in the limit of $l \to 0$, and low temperatures}

\begin{figure}
\centering
\resizebox{0.9\columnwidth}{!}{
\includegraphics{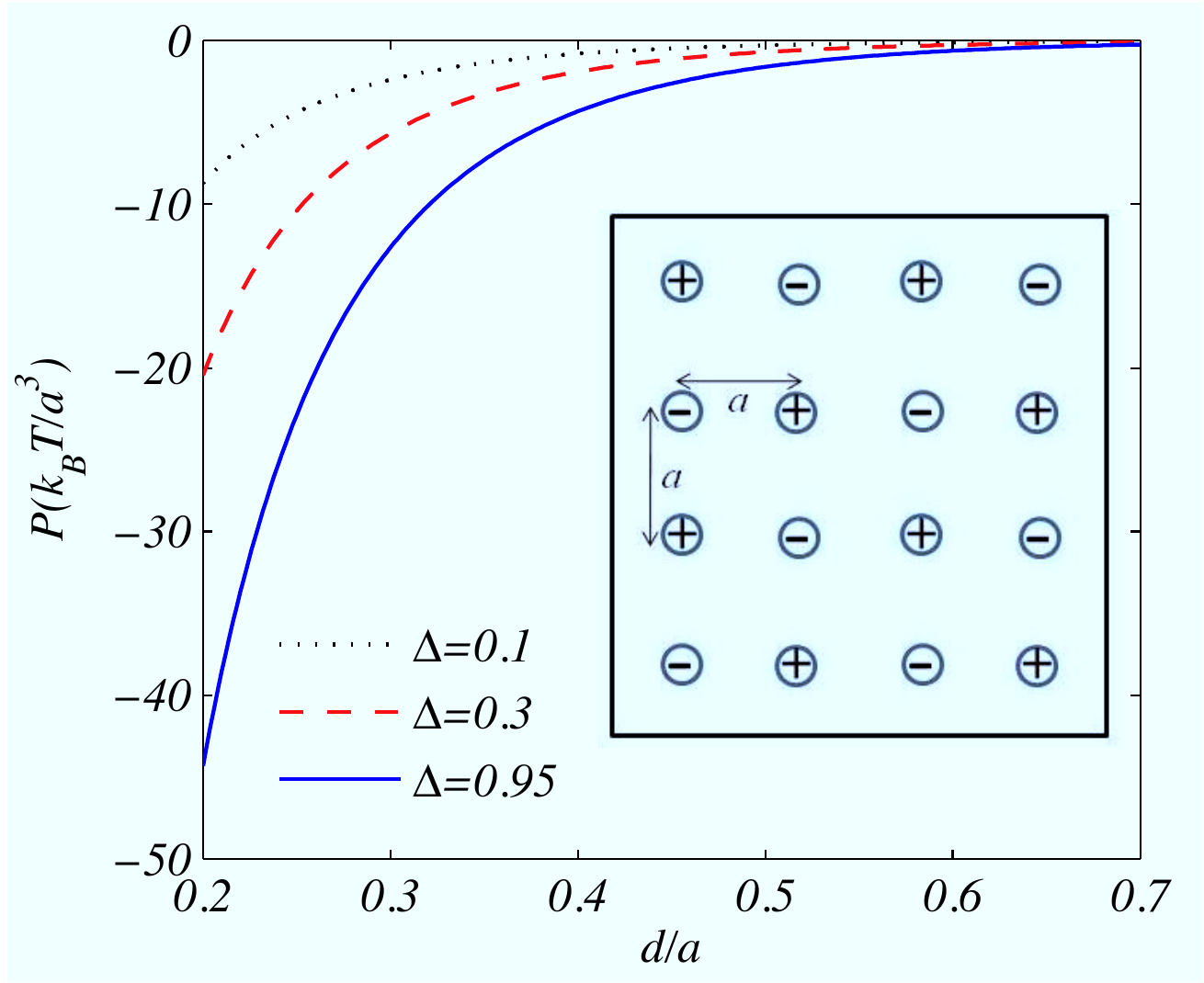}}
\caption{(Color online) The pressure, $P$, acting on each slab as a function of the separation between them, $d$. Inset: The pattern of the charges of each slab for divalent counterions in the limit of $l \to 0$ and at low temperatures (see the text). }
\label{fig:P-d}
\end{figure}

As one can see in Figs. \ref{fig:2slab-density-l} and \ref{fig:2slab-zeq}, for very small values of $l$ the counterions prefer to locate very close to the slabs. Therefore when the temperature is low, the entropy is negligible and half of the counterions are distributed in the vicinity of each slab. Since in this limit the electrostatic interaction is dominant, the counterions follow the pattern of the slabs surface charge and make a 2D lattice to minimize the free energy (or the dominant electrostatic energy). In this case the valence of the counterions is important. For example for mono-valent counterions, the 2D lattice of the counterions is exactly the same as the lattice of the charged regions on the slabs, whereas the lattice constant for the case of divalent counterions is $\sqrt 2 a$. In this section, we consider divalent counterions and find the interaction between the two slabs and investigate the effect of the dielectric inhomogeneity. Regarding the large energy barrier against the counterions movement out of their equilibrium positions and small width of the energy wells, when the temperature is low, one can imagine that the counterions cannot leave their equilibrium positions and their fluctuations around their equilibrium positions is negligible. In this condition, the problem is equivalent to two charged slabs, with new effective charge density, with no counterions between them. Due to the location of the divalent counterions on each slab, the effective pattern of charges of each slab follows a periodic positive/negative scheme on a 2D lattice of spacing $a$, as shown in the inset of Fig. \ref{fig:P-d}. Therefore, the charge density of each slab can be written as
\begin{eqnarray}
\rho_{slab1}=e\delta(z+D) \times {} \nonumber \\
&& {}\hspace{-2cm}\sum_{m,n=-\infty}^{+\infty}(-1)^{m+n}\delta(x-ma) \delta(y-na) {} \nonumber \\
&& {}\hspace{-3.6cm} \rho_{slab2}=e\delta(z-D) \times {} \nonumber \\
&& {}\hspace{-3.4cm} \sum_{m,n=-\infty}^{+\infty}(-1)^{m+n}\delta(x-ma-\phi_{x})
\delta(y-na-\phi_{y}), \nonumber
\\  \label{eq:rho-2slabs}
\end{eqnarray}
where $\phi_x$ and $\phi_y$ are the relative displacement of the two slabs in the $x$ and $y$ directions. Using this charge distribution and the Green's function of the system, Eq. (\ref{green}), the electric potential can be calculated. Substituting this potential into the relation of energy, $U=\frac{1}{2}\int\rho\phi dV$, the total energy per lattice unit is found as
\begin{eqnarray}
U_{tot}=\frac{e^2}{2\varepsilon_{2}a^2}\sum_{m,n=0}^{+\infty}\frac{(1+W(kb))^2}{k(e^{4kD}-W^2(kb))}\times
\nonumber \\
&& \hspace{-7.5cm}\left(W(kb)+e^{2kD} \cos \left[ \frac{(2n-1)\pi}{a}\phi_{x} \right] \cos \left[\frac{(2m-1)\pi}{a}\phi_{y}\right] \right),
\nonumber \\
\label{eq:Utot-2slab}
\end{eqnarray}
where $k=\frac{\pi}{a}\sqrt{\left(2n-1\right)^2+\left(2m-1\right)^2}$. $\phi_x$ and $\phi_y$ are chosen to minimize the above energy $\cos \left[ \frac{\left(2n-1\right)\pi}{a}\phi_x \right] \cos \left[\frac{\left(2n-1\right)\pi}{a}\phi_y \right] = -1$ and the total energy per lattice unit becomes
\begin{eqnarray}
U_{tot}=-\frac{e^2}{2\varepsilon_{2}a^2}\sum_{m,n=1}^{+\infty}
\frac{(1+W(kb))^2}{k\left(e^{2kD}-W(kb)\right)}. \label{eq:Utot}
\end{eqnarray}

Using the relation $P=-\frac{1}{2 a^2} \frac{\partial U_{tot}}{\partial D}$ the pressure acting on each slab can be found as
\begin{eqnarray}
P=-\frac{e^2}{2\varepsilon_{2}a^4}\sum_{m,n=1}^{+\infty}
\frac{\left[1+W(kb)\right]^2 e^{2kD}}{\left[e^{2kD}-W(kb)\right]^2}. \label{eq:P-2slab}
\end{eqnarray}
We see that the pressure is always negative meaning that the interaction is attractive. In Fig. \ref{fig:P-d}, the dependence of pressure, $P$, on the separation between the two slabs, $d$, is shown for different values of $\Delta$. One can see that by increasing the value of $\Delta$, the attraction between the slabs becomes stronger.

\subsubsection{Interaction between the two slabs in the SC regime at finite temperatures}

At room temperature, the total free energy of the system consists of the interaction energy of the two slabs with each other and the counterions free energy. The two slabs interaction energy can be obtained using the Green function and the slabs charge distribution as
\begin{eqnarray}
\frac{U_{slabs}}{Nk_{B}T}=-\frac{\pi Z^2l_{B}D}{a^2}+\nonumber \hspace{+5.1cm}\\ \frac{2\pi Z^2 l_{B}}{a^2}\sum_{n=1}^{\infty}\sum_{m=1}^{\infty}
\frac{(1+W(kb))^2}{k(e^{2kD}-W(kb))}\left(\frac{\sin k_{x}l/2\sin k_{y}l/2}{k_{x}k_{y}l^2/4}\right)^2  \nonumber\\
&& \hspace{-9.3cm}+\frac{2\pi Z^2 l_{B}}{a^2}\sum_{n=1}^{\infty}\frac{(1+W(k'b))^2}{k'(e^{2k'D}-W(k'b))}\left(\frac{\sin k'l/2}{k'l/2}\right)^2
\end{eqnarray}
In the SC limit, the counterions are frozen in a 2D lattice due to their strong interactions \cite{SC}. In this regime it can be assumed that the counterions are positioned on a surface parallel to the slabs at the coordinates of $x=na$ and $y=ma$ and can only move in the $z$ direction. With these considerations, the system total free energy can be written as
\begin{equation}
\frac{F}{Nk_{B}T}=\frac{U_{slabs}}{Nk_{B}T}-\ln\int_{0}^{d}dz\exp\left(-\frac{E}{k_{B}T}\right), \label{eq:F-2slab}
\end{equation}
where $E$ denotes the electrostatic energy of a counterion that is given by Eq. (\ref{eq:2slab-energy}). The pressure acting on a slab can be obtained using the relation $P=-\frac{1}{2a^2} \frac{\partial F}{\partial D}$.

\begin{figure}
\centering
\resizebox{0.9\columnwidth}{!}{
\includegraphics{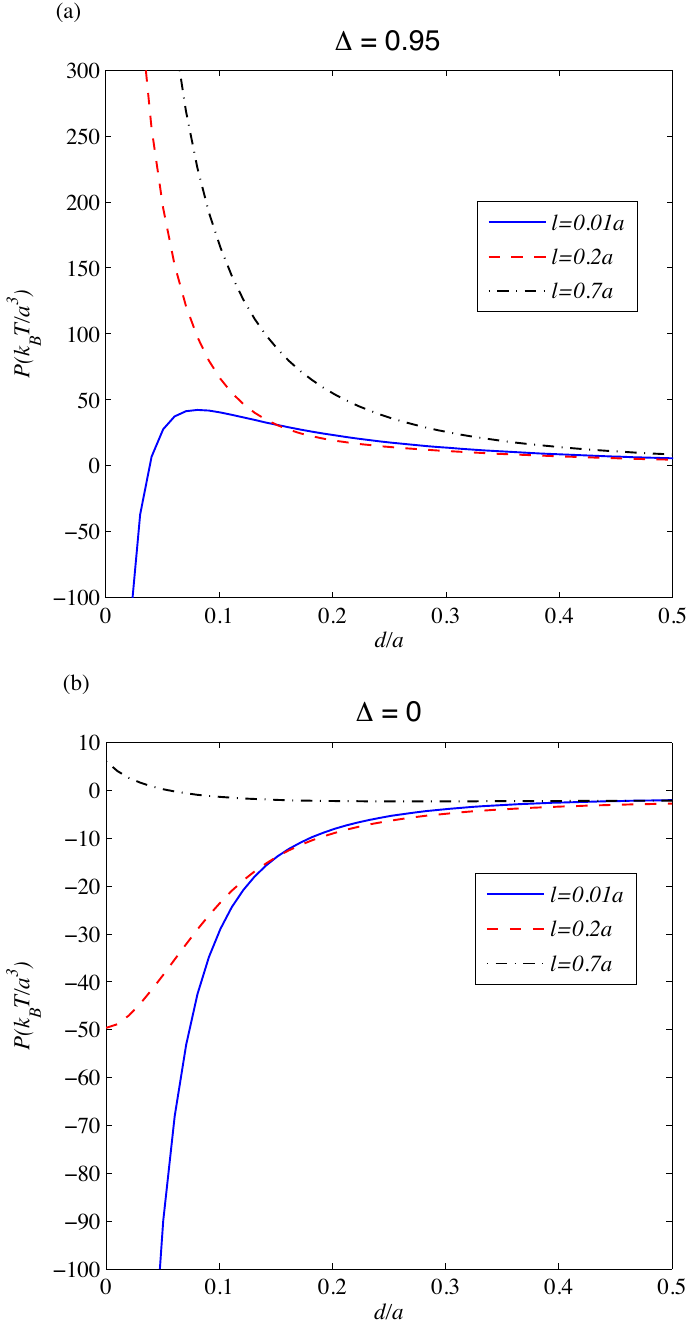}}
\caption{(Color online) The force per unit area on a slab as a function of $d$, the distance between the two slabs, for different values of $l$. Figures (a) and (b) correspond to $\Delta=0.95$ and $\Delta=0$, respectively.}
\label{fig:P-d-2}
\end{figure}

In Fig. \ref{fig:P-d-2}(a), the dependence of $P$, the force per unit area, is shown as a function of distance between the two slabs, $d$, for $\Delta=0.95$ and different values of $l/a$. Three interactions determine the total force on a slab: The counterions interaction with the physical charges on the slabs, image-counterion interaction and the interaction between the physical charges of the two slabs. As one can see easily, the first force is always attractive, whereas the second and the third ones are repulsive. When the slabs are far from each other, the first force is always dominant and the total force is attractive. In closer separations, the effects arisen from the dielectric inhomogeneity become important and the force between the slabs becomes repulsive. Hence, for larger values of $l$, an equilibrium separation between the slabs exists. This is in agreement with the result of Ref. \cite{SC-dielectric1} for uniform distribution of charge on the slabs. But for smaller values of $l$, when the slabs are in close separation, the electrostatic attraction dominates again and the slabs completely attract each other after passing a repulsion barrier. It is interesting to note that the latter behavior is completely different from the behavior of the slabs with uniform surface charge which shows the important role of the charge discreteness.

 \begin{figure*}
 \centering
\resizebox{1.99\columnwidth}{!}{
\includegraphics{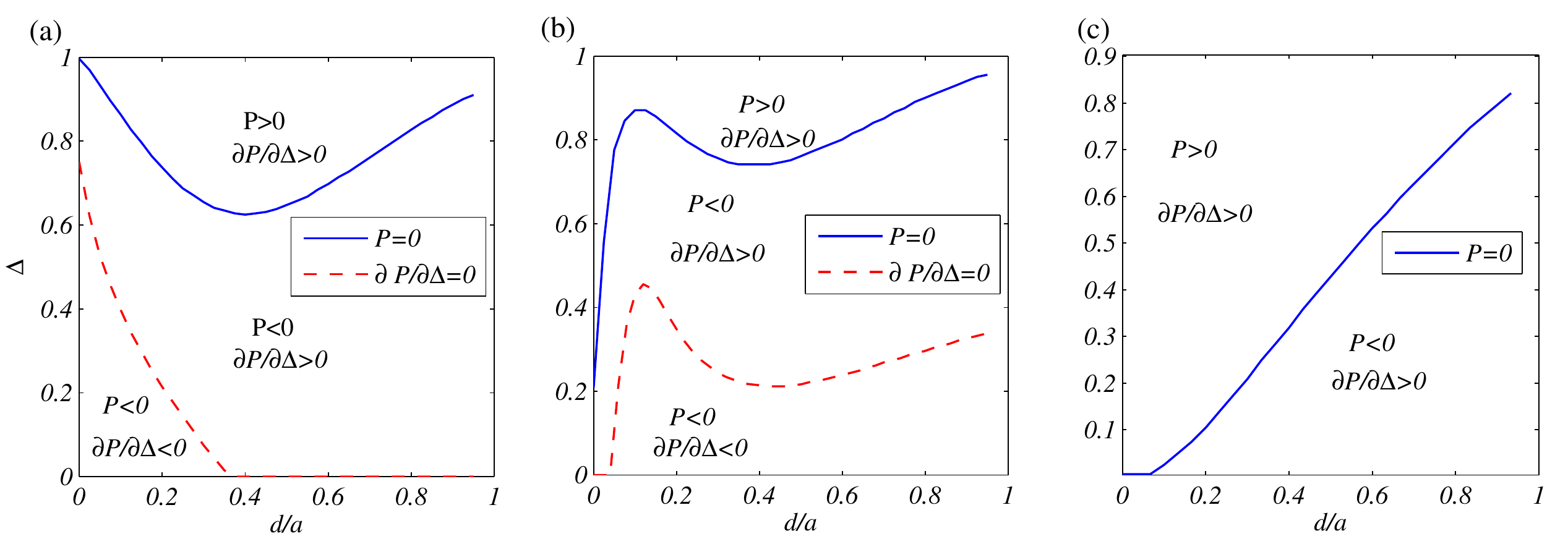}}
\caption{(Color online) The phase space of the system for three representative cases of (a) $l = 0.01 a$, (b) $l=0.15 a$, and (c) $l = 0.7 a$. The plots are delineating the different regimes in the parameter space of two charged slabs, where $d$ and $\Delta$ denote the distance between the two slabs and dielectric inhomogeneity, respectively. The solid line corresponds to $P=0$ and the dashed line corresponds to $\partial P/ \partial \Delta = 0$.  }
\label{fig:phase}
\end{figure*}

The behavior of $P$ is shown in Fig. \ref{fig:P-d-2}(b) as a function of $d$ for $\Delta=0$ and different values of $l/a$. As one can see, when $l$ is large and the charge distribution on the slabs is approximately uniform, the force on the slabs in close separations is repulsive which is the result of the counterions entropy \cite{SC-uniform}. The interaction force becomes attractive for large separations and there is an equilibrium distance for the two slabs. Furthermore, for small values of $l$, the electrostatic attraction becomes stronger and overcomes the repulsion. As it can be seen in this figure, at small values of $l$ the two slabs attract each other in all separations. This is in agreement with the result of Ref. \cite{SC-discrete2} that the charge discreteness strengthens the attraction between the slabs.

In Figs. \ref{fig:phase}(a)--(c), attraction and repulsion regions in $\Delta$-$d$ plane are shown for the system. One can see that by increasing the charge discreteness, the structure of the phase space changes completely. By decreasing $l$, the repulsion region in small separations disappears and a repulsion region appears in the intermediate separations. At each value of $l$, with increasing $\Delta$ the repulsion region grows. The interesting point in these plots is that for specific values of $l$ and $\Delta$, there are two equilibrium separations between the two slabs. 

\begin{figure}
\centering
\resizebox{0.99\columnwidth}{!}{
\includegraphics{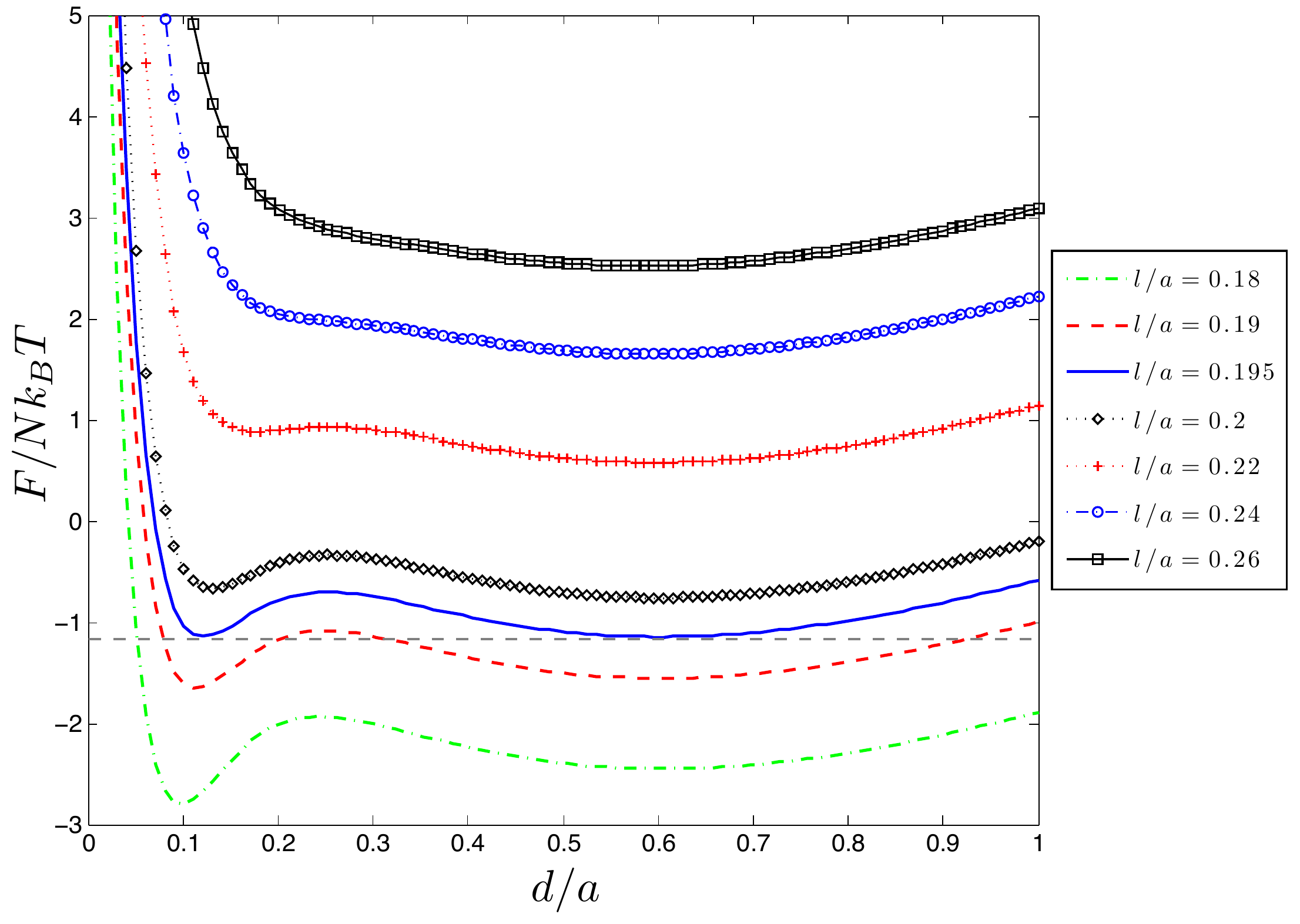}}
\caption{(Color online) Free energy of the system for $\Delta=0.8$ and several values of $l$. At $l/a \simeq 0.195$, two minima of the free energy have the same value (the horizontal dashed line).  }
\label{fig:freeenergy}
\end{figure}

In Fig. \ref{fig:freeenergy}, the behavior of the free energy of the system is shown as a function of $d/a$ for different values of $l/a$ and $\Delta = 0.8$. As one can see, for the values of $l/a \gtrsim 0.25$, the free energy has two minima showing two equilibrium separations between the two slabs. It is interesting to note that these two minima have the same free energy for $l/a \simeq 0.195$. For $l/a \lesssim 0.195$, the free energy has a global minimum when the distance of the two slabs is about $d/a<0.2$. When $l/a > 0.195$, the global minimum of the free energy tends to the larger distance of the two slabs, say $d/a>0.5$. As it can be seen in the figure, for the larger values of $l/a$ (e.g. $l/a \gtrsim 0.26$), the free energy has only one minimum that is located in $d/a \gtrsim 0.5$.

\section{Discussion and concluding remarks}

Strongly-charged biological macromolecules such as DNA and filamentous actin are in the limit of strong coupling in biological conditions. For these macromolecules in solution, it is known that multivalent counterions are mostly distributed in their vicinity. When the small ions are very close to these macromolecules or when two of them are very close to each other, the curvature of their surface can be ignored in calculation of the electrostatic interactions. Therefore, our study of the interaction between two flat dielectric slabs could be useful in enlightening the phenomenon of interaction between biological charged macromolecules in the presence of multivalent counterions. As examples, like-charge attraction between actin filaments \cite{Wong-2003}, the interaction between DNA and charged proteins \cite{protein-DNA-ligand} and a system of proteins confined in a polyelectrolyte brush \cite{Ballauff} can be considered. 
In most of computer simulation studies of such systems dielectric inhomogeneity and real charge distribution on the surface of the macromolecules have not been taken into account. This is because of intrinsic difficulty of calculation of the electrostatic interactions in the presence of dielectric inhomogeneities and massive calculations needed for modeling the real charge distribution on the macromolecules. In this subject only simple systems like single or a couple of dielectric slabs with uniform charge distribution on their surfaces have been studied.

By calculation of the Green function for two dielectric slabs (the geometry shown in Fig. \ref{fig1}), we calculated the electrostatic interaction between them taking into consideration both the dielectric inhomogeneity and the charge discreteness.
In the strong coupling limit, it is found that the amount of dielectric constant difference between the slabs and the environment and the discreteness of charge on the slabs have opposing effects on the equilibrium distribution of the counterions between the slabs. Increasing the amount of dielectric constant difference increases the tendency of the counterions toward the middle of the space between the slabs. Discreteness of charge on the slabs however, pushes the counterions to the surface of the slabs. At low temperatures, the interaction between the slabs is attractive, the equilibrium separation between the slabs vanishes and the strength of their attraction increases with increasing the amount of the dielectric constant difference. At room temperature, the slabs may attract each other and come together, stay in an equilibrium separation or have two equilibrium separations with a barrier in between depending on the system parameters.

Our results showed that for a system of dielectrics of a simple geometry, namely two dielectric slabs, taking into consideration the dielectric inhomogeneity and leaving the reality of discreteness of charge on the dielectrics is not a valid approach when they are close to each other. In the extreme of point charges on the dielectric slabs, the effect of charge discreteness completely dominates over the effect of dielectric inhomogeneity. For example, distribution of counterions in the middle of the two slabs because of the dielectric inhomogeneity (as described in Ref. \cite{SC-dielectric1}) is the effect of the assumption of uniform charge distribution.  Our results show that by taking into account the reality of charge discreteness on the slabs, all the counterions distribute in the vicinity of the slabs surfaces and the middle of the slabs correspond to the minimum of the counterions density.

When the surface charge density is uniform, the pressure between the plates is repulsive at small separations and the plates stand in an equilibrium distance from each other. This repulsion is due to the counterions entropy and the dielectric inhomogeneity. As the charge discreteness increases, the repulsion at close separations disappears and the plates completely attract each other after passing a repulsion barrier in the intermediate separations. Also, dielectric inhomogeneity widens the repulsion region between the slabs and increases the repulsion strength. For highly discretized surface charge, the interaction between the slabs was shown that is attractive at small inter-slab separations. In this case, the dielectric inhomogeneity may increase or decrease the attractive pressure, depending on the value of the parameters.

Another point to note is the large depth of the energy wells in Figs. \ref{fig:1slab-energy} and \ref{fig:1slab-E-x} relative to the thermal energy, $k_BT$. It shows that the charge discreteness and the dielectric inhomogeneity strongly affect the equilibrium distribution of the counterions and hence the interaction between the slabs. Regarding the competing effects of charge discreteness and the dielectric inhomogeneity, our results showed that considering only one of these realities and leaving out the other one may crucially bias the results of the theory. 

In this paper, distribution of the counterions and interaction between the slabs are obtained from the SC theory at the leading order of the expansion. It has been shown that this approximation is correct for discretely charged surfaces in coupling parameters larger than 20 \cite{SC-discrete2}. Many biological systems lie in this range. However, when the separation between the two charged plates gets very large, the first order term becomes smaller than the second term and this approximation fails. This critical distance is found for uniformly charged plates with and without dielectric inhomogeneity. Computer simulations have also been used to find the validity range of the SC theory in these systems \cite{SC,SC-dielectric1}. 
As a result, computer simulations considering both the dielectric inhomogeneity and the charge discreteness effects seem essential.

The SC theory is obtained from an expansion of the system parameters in inverse powers of the coupling parameter $\Xi$. Theoretically, SC formalism is exact in the limit of $\Xi\to\infty$, however computer simulations have shown that the results become accurate for values of $\Xi \gtrsim 100$  \cite{SC}. In biological conditions, $\Xi=100$  is equivalent to $a\simeq 1nm$ in our model or surface charge density $\sigma_S \simeq 1\frac{e}{nm^2}$ which is of the order of the typical charge density of charged biological membranes and DNA \cite{SC-uniform}. For discretely charged surfaces, SC theory has been shown to be exact even in lower values of $\Xi$ (down to $\Xi\simeq20$) due to the weakened correlations between the counterions \cite{SC-discrete2}. Also, it has been shown that the presence of the dielectric inhomogeneity widens the validity range of the SC theory \cite{SC-dielectric1}.

The validity range of the SC theory in the case of a single charged plate is $z < a_{\perp}$ in which $z$ is the separation between counterions and the surface and $a_{\perp}$ is the mutual separation between the counterions parallel to the surface. This range for the system of two parallel charged surfaces is $d < a_{\perp}$ in which $d$ is the separation between the surfaces \cite{SC}. In our model, the lattice spacing of the regular charge distribution on the surfaces, $a$, plays the role of $a_{\perp}$ of the SC theory. Considering that the most of our results have been obtained using parameters in the range of $d < a$, the validity of the SC theory is preserved. 

One should note that a sharp boundary is assumed between inside and outside of the slabs in our model. In reality, it is known that such a sharp boundary is not the case and an improvement of our model could be the consideration of a smooth profile of the dielectric constant in the slabs boundaries.
Some additional points in the way of approving the theory are as follows. The grand canonical ensemble is more proper for study of the real systems such as the system of dielectric slabs with counterions studied here but the SC theory is developed in the canonical ensemble. Developing a theory in which the number of the counterions between the slabs is not constant seems as a valuable step. The finite distance of the fixed charges from the slab surface \cite{SC-discrete1,SC-discrete2}, random distribution of the charged domains \cite{rand}, displacement of the charged domains by changing the system parameters in some cases \cite{displ} and possibility of their hydration \cite{hydr} should be considered in development of the theory. Also, experimental studies on the effects of different factors on the distribution of the counterions and interaction between charged dielectrics can also be insightful. For example, the effect of the dielectric inhomogeneity on the interaction of two charged dielectrics seems to be possible by changing the solvent \cite{CI} and is not studied yet to the authors knowledge.

\section{ACKNOWLEDGMENT}
We are grateful to F. Julicher, S. N. Rasuli, and M. F. Miri for interesting discussions and comments.

\end{document}